\documentclass[twoside,twocolumn,9pt]{article}

\usepackage{extsizes}
\usepackage[super,sort&compress,comma]{natbib} 
\usepackage[version=3]{mhchem}
\usepackage[left=1.5cm, right=1.5cm, top=1.785cm, bottom=2.0cm]{geometry}
\usepackage{balance}
\usepackage{mathptmx}
\usepackage{sectsty}
\usepackage{graphicx} 
\usepackage{lastpage}
\usepackage[format=plain,justification=justified,singlelinecheck=false,font={stretch=1.125,small,sf},labelfont=bf,labelsep=space]{caption}
\usepackage{float}
\usepackage{fancyhdr}
\usepackage{fnpos}
\usepackage[english]{babel}
\addto{\captionsenglish}{%
  \renewcommand{\refname}{Notes and references}
}
\usepackage{array}
\usepackage{droidsans}
\usepackage{charter}
\usepackage[T1]{fontenc}
\usepackage[usenames,dvipsnames]{xcolor}
\usepackage{setspace}
\usepackage[compact]{titlesec}
\usepackage{hyperref}

\usepackage{braket}

\usepackage{epstopdf}

\definecolor{cream}{RGB}{222,217,201}

\begin{document}

\pagestyle{fancy}
\thispagestyle{plain}
\fancypagestyle{plain}{
\renewcommand{\headrulewidth}{0pt}
}

\makeFNbottom
\makeatletter
\renewcommand\LARGE{\@setfontsize\LARGE{15pt}{17}}
\renewcommand\Large{\@setfontsize\Large{12pt}{14}}
\renewcommand\large{\@setfontsize\large{10pt}{12}}
\renewcommand\footnotesize{\@setfontsize\footnotesize{7pt}{10}}
\makeatother

\renewcommand{\thefootnote}{\fnsymbol{footnote}}
\renewcommand\footnoterule{\vspace*{1pt}%
\color{cream}\hrule width 3.5in height 0.4pt \color{black}\vspace*{5pt}} 
\setcounter{secnumdepth}{5}

\makeatletter 
\renewcommand\@biblabel[1]{#1}            
\renewcommand\@makefntext[1]%
{\noindent\makebox[0pt][r]{\@thefnmark\,}#1}
\makeatother 
\renewcommand{\figurename}{\small{Fig.}~}
\sectionfont{\sffamily\Large}
\subsectionfont{\normalsize}
\subsubsectionfont{\bf}
\setstretch{1.125} 
\setlength{\skip\footins}{0.8cm}
\setlength{\footnotesep}{0.25cm}
\setlength{\jot}{10pt}
\titlespacing*{\section}{0pt}{4pt}{4pt}
\titlespacing*{\subsection}{0pt}{15pt}{1pt}

\fancyfoot{}
\fancyfoot[LO,RE]{\vspace{-7.1pt}\includegraphics[height=9pt]{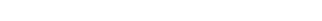}}
\fancyfoot[CO]{\vspace{-7.1pt}\hspace{11.9cm}\includegraphics{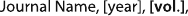}}
\fancyfoot[CE]{\vspace{-7.2pt}\hspace{-13.2cm}\includegraphics{head_foot/RF}}
\fancyfoot[RO]{\footnotesize{\sffamily{1--\pageref{LastPage} ~\textbar  \hspace{2pt}\thepage}}}
\fancyfoot[LE]{\footnotesize{\sffamily{\thepage~\textbar\hspace{4.65cm} 1--\pageref{LastPage}}}}
\fancyhead{}
\renewcommand{\headrulewidth}{0pt} 
\renewcommand{\footrulewidth}{0pt}
\setlength{\arrayrulewidth}{1pt}
\setlength{\columnsep}{6.5mm}
\setlength\bibsep{1pt}

\makeatletter 
\newlength{\figrulesep} 
\setlength{\figrulesep}{0.5\textfloatsep} 

\newcommand{\topfigrule}{\vspace*{-1pt}%
\noindent{\color{cream}\rule[-\figrulesep]{\columnwidth}{1.5pt}} }

\newcommand{\botfigrule}{\vspace*{-2pt}%
\noindent{\color{cream}\rule[\figrulesep]{\columnwidth}{1.5pt}} }

\newcommand{\dblfigrule}{\vspace*{-1pt}%
\noindent{\color{cream}\rule[-\figrulesep]{\textwidth}{1.5pt}} }

\makeatother

\twocolumn[
  \begin{@twocolumnfalse}
{
}\par
\vspace{1em}
\sffamily
\begin{tabular}{m{4.5cm} p{13.5cm} }

\includegraphics{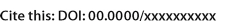} & \noindent\LARGE{\textbf{Coherent state switching using vibrational polaritons in an asymmetric double-well potential}} \\
\vspace{0.3cm} & \vspace{0.3cm} \\

 & \noindent\large{Lo\"ise Attal,\textit{$^{a}$} Florent Calvo\textit{$^{b}$} Cyril Falvo\textit{$^{a,b}$} and Pascal Parneix,\textit{$^{a,\dag}$} } \\



\includegraphics{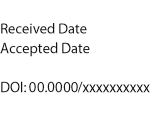} & \noindent\normalsize{
The quantum dynamics of vibrational polaritonic states arising from the interaction of a bistable molecule with the quantized mode of a Fabry-Perot microcavity is investigated using {a generic} asymmetric double-well potential as a simplified one-dimensional model of a reactive molecule. After discussing the role of the light-matter coupling strength in the emergence of avoided crossings between polaritonic states, we investigate the possibility of using these crossings to trigger a dynamical switching of these states from one potential well to the other. Two schemes are proposed to achieve this coherent state switching, either by preparing the molecule in an appropriate vibrational excited state before inserting it into the cavity, or by applying a short laser pulse inside the cavity to obtain a coherent superposition of polaritonic states. The respective influences of the dipole amplitude and potential asymmetry on the coherent switching process are also discussed.
}


\end{tabular}

\end{@twocolumnfalse} \vspace{0.6cm}]


\renewcommand*\rmdefault{bch}\normalfont\upshape
\rmfamily
\section*{}
\vspace{-1cm}


\footnotetext{\textit{$^{a}$~Université Paris-Saclay, CNRS, Institut des Sciences Moléculaires d'Orsay, 91405 Orsay, France}}
\footnotetext{\textit{$^{b}$~Université Grenoble Alpes, CNRS, LIPhy, 38000 Grenoble, France }}

\footnotetext{\dag~pascal.parneix@universite-paris-saclay.fr}



\section{\label{sec:Intro}Introduction}

{Over the recent years, confining molecular systems in small
Fabry-Perot (FP) cavities has been suggested as a way to control their physical and chemical properties.\cite{Schwartz2011,Hutchison2012,Wang2014,Long2015,Simpkins2015,Flick2018,Ribeiro2018,Thomas2019,Li2021a,Li2021b}
Polaritonic chemistry, which exploits the quantization of the electromagnetic field inside such cavities and the resulting hybrid
states that arise due to strong light-matter interactions, was
first investigated using electronic
transitions\cite{Schwartz2011,Hutchison2012,Wang2014} before being
extended to vibrational degrees of
freedom.\cite{Shalabney2015,Long2015,Simpkins2015,Thomas2016,Thomas2019,Takele2020,Wright2023}
The so-called polaritonic states could in principle enable a greater
control over chemical processes such as photoisomerization or
reactivity, provided that entanglement of the light-matter degrees of
freedom survives on time scales greater than both the photon lifetime
in the cavity and the molecular coherence time.}

{So far, the changes in the molecular properties caused by vibrational
strong coupling (VSC) in a cavity have been probed using various
spectroscopy methods, including linear infrared (IR)
spectroscopy,\cite{George2015,Wright2023} Raman
spectroscopy\cite{Shalabney2015,Pino2015,Strashko2016,Ahn2021} as well
as 2D-IR spectroscopy.\cite{Saurabh2016,Cohn2022,Xiong2023} Among the
various properties that could be altered under VSC, energy transport,
photochemical reactivity and thermally driven ground-state chemical
reactions have been scrutinized.\cite{George2015,George2016,Muallem2016,Vergauwe2016,Kapon2017,Takahashi2019}
However, the difficulties inherent to these experiments\cite{simpkins21}
 have raised a number of controversies regarding their reproducibility\cite{imperatore21,wiesehan21} and interpretation.\cite{simpkins21,thomas23}}

{From a general perspective,} and since the vacuum field amplitude $E_0$
of a confined light mode with frequency $\omega_c$ and mode volume
$V_c$ is proportional to $1/\sqrt{V_c}$, strong coupling requires the
FP cavity to be small, \textit{e.g.} electronic polaritons typically
require nanoscale cavities. For vibrational polaritons, since
transitions occur in a prescribed electronic state, $\omega_c$
typically lies in the range of 10$^2$--10$^3$ cm$^{-1}$ leading to a
cavity length $L\approx 1$--10~$\mu$m. As shown by the Tavis-Cummings
model,\cite{Tavis1968} the light-matter coupling strength depends on
the number $N$ of molecules in the cavity and varies as $\sqrt{N}$,
leading to the use of dense media in most experimental studies so
far. However, the formation of rovibrational polaritons of gaseous
methane confined in an FP microcavity was very recently
suggested.\cite{Wright2023} {Besides allowing larger cavity volumes, diluted media enable much longer photon lifetimes, in excess of the nanosecond for the aforementioned work on methane.}


The experimental results on vibrational polaritons have opened a
number of fundamental questions that stimulated the interest of the
theoretical
community,\cite{Groenhof2018,Ribeiro2018a,Hernandez2019,Triana2020,Herrera2020,Fischer2020,Triana2021,Li2021,Wang2021,Fischer2022a,Mandal2022,Gomez2023}
aimed at clarifying the role of dissipative effects in the cavity,
the importance of dark states and the role of relaxation processes
such as intramolecular vibrational relaxation. For {single} molecules, vibrational polaritons have been extensively
studied using model potentials such as Morse potentials
\cite{Groenhof2018,Hernandez2019,Triana2020,Fischer2020,Triana2021,Li2021}
or symmetric double-well
potentials.\cite{Fischer2020,Fischer2022b} Rovibrational polaritons
have also been explored by including two cavity modes with orthogonal
polarisation,\cite{Fischer2022a} and particular interest for the
possibilities offered by multi-mode cavities have recently
emerged.\cite{George2016, Balasubrahmaniyam2021,Mandal2023} The case
of an assembly of confined molecules interacting with each other was
also studied, and the importance of dark states on the properties of
the polaritonic states was specifically
emphasized.\cite{Ulusoy2019,Herrera2020,Mandal2022,Gomez2023} However,
and with the exception of very recent efforts 
{to treat polyatomic molecules fully coupled to the cavity at an \textit{ab initio} level using cavity Born-Oppenheimer models,}\cite{Schnappinger:2023aa,Schnappinger:2023ab} most
theoretical studies have relied on quite simplified and low
dimensional models of the molecular systems used in experiments. Such
approaches notably struggle to take into account the effects of the
environment and the interaction mechanisms between polaritons.%


In the present work, {and building upon earlier efforts from Triana, Hernandez and coworkers,\cite{Hernandez2019,Triana2020,Triana2021}} we consider {a generic} one-dimensional asymmetric double-well potential as a simplified model of the electronic ground state of a molecule undergoing a chemical reaction. 
Under sufficient coupling between the vibrational mode of the molecule and the electromagnetic mode of the cavity, stationary vibrational polaritonic states are formed. 
These hybrid light-matter states exhibit avoided crossings at specific values of the coupling strength that will be used to trigger coherent state switching between the two wells of the asymmetric potential. The localization of some polaritonic states in one potential well or the other is made possible by the asymmetry of the potential, which lifts the symmetry constraints preventing the localization of the eigenstates in symmetric potentials.
Different ways of altering the polaritonic states to induce coherent switching are considered, either by preparing the molecule in a specific vibrational state before inserting it into the cavity or by exposing the hybrid system to an external {picosecond} laser pulse. Efficient coherent state switching between vibrational polaritons located in each potential well is numerically observed for suitable values of the light-matter coupling.

In section \ref{sec:Theory}, the theoretical framework used to describe the one-dimensional model system and its interaction with the microcavity is presented, and some computational details are provided. The computational results are presented and discussed in section \ref{sec:Results}, first for stationary states, and then under the two out-of-equilibrium aforementioned conditions. Section \ref{sec:Conclusion} concludes the article by summarizing the main results and opening some perspectives for future efforts.

\section{Theory}
\label{sec:Theory}


\subsection{Hamiltonian}
\label{ssec:Theory}

We consider a single molecule ($N = 1$) described by a simple
one-dimensional model with coordinate $q$ and momentum $p$, in
vibrational strong coupling with a single-mode FP microcavity. The
Pauli-Fierz Hamiltonian\cite{PF1938} describing such a molecule
interacting with the quantized electromagnetic mode of the cavity can
be given by
\begin{equation}
   \tilde{H} = \tilde{H}_{{\rm{cav}}} + \tilde{H}_{{\rm{mol}}} +
   \tilde{H}_{{\rm{int}}},
\end{equation}
in which $\tilde{H}_{{\rm{cav}}}$, $\tilde{H}_{{\rm{mol}}}$ and
$\tilde{H}_{{\rm{int}}}$ denote the contributions of the cavity, of
the molecule, and of their interaction, respectively.  Without loss of
generality, the vibrational Hamiltonian of the molecule is given by
\begin{equation}
    \tilde{H}_{{\rm{mol}}}(q,p) = \frac{p^2}{2 \mu} + \tilde{V}(q),
\end{equation}
with $\mu$ the reduced mass of the molecule and $\tilde{V}(q)$ its
potential energy operator. This Hamiltonian can be diagonalized to
obtain the free molecule vibrational eigenstates $\ket{v}$ and their
associated energies $E(v) = \omega_v$.

The single-mode cavity Hamiltonian, which is characterized by its mode
frequency $\omega_c$, can be written using the creation and
annihilation operators as
\begin{equation}
\tilde{H}_{{\rm{cav}}} = \omega_c \, \tilde{a}^\dagger \tilde{a},
\end{equation}
in which the factor $\hbar$ was omitted, as in all equations to
follow.  The cavity mode frequency is chosen to be in resonance with
the fundamental molecular transition to ensure strong light-matter
interaction.
%
This interaction depends on the vacuum field amplitude $E_0 = \sqrt{
  \omega_c/V_c\epsilon_0}$ and on the dipole moment $\tilde{d}$ of the
molecule.  In the following, the rotation of the molecule is
neglected, and the dipole moment vector is assumed to be aligned with
the polarization vector of the cavity mode. This allows to only
consider scalar quantities in the coupling Hamiltonian. By taking into
account both the dipolar interaction and dipole self-energy
contribution,\cite{Fischer2020} the molecule-light interaction can be
written as
\begin{equation}
    \tilde{H}_{{\rm{int}}} = E_0 \tilde{d} (\tilde{a}^\dagger +
    \tilde{a}) + \frac{E_0^2}{ \omega_c} \tilde{d}^2.
\end{equation}
The strength of this coupling can be characterized by the
dimensionless parameter $\lambda_c$ given by
\begin{equation}
\lambda_c= \frac{E_0 \,  |  d_{01} | }{ \omega_c},
\label{Eq:lambda}
\end{equation}
with $d_{01}=\langle 0 | \tilde{d} | 1 \rangle$ the dipole moment
between the ground and the first vibrational excited state of the
molecule.

Using the vibrational eigenstates $\ket{v}$ of the free molecule and
the cavity mode Fock states $\ket{n}$, the Pauli-Fierz Hamiltonian can
be expressed in the uncoupled basis set $ | v,n\rangle = | v\rangle
\otimes | n\rangle$ as
\begin{align}
\tilde{H} &= \sum_n\sum_v \left(n \omega_c \ + \omega_v \right)
\ket{v,n} \bra{v,n}\nonumber \\ &+ \lambda_c \, \omega_c \sum_n \sum_v
\sum_{v'} \frac{ \langle v' | \tilde{d} | v \rangle}{ | d_{01} | } \,
\left(| v', n+1\rangle \langle v,n | + | v', n\rangle \langle v, n+1 |
\right) \nonumber \\ &+ \lambda_c^2 \, \omega_c \sum_n\sum_v \sum_{v'}
\frac{\langle v' |\tilde{d}^2 | v \rangle }{d_{01}^2} \, | v',n\rangle
\langle v,n |,
 \label{Eq:H_vn}
\end{align}
where Eq.~(\ref{Eq:lambda}) was used to introduce the coupling
parameter $\lambda_c$ in the Hamiltonian.

\subsection{Asymmetric double-well potential}

\begin{table}[b]
\small
  \caption{Parameters chosen for the potential energy and dipole
    moment functions. The numbers $N_\ell$ and $N_r$ of vibrational
    states localized in the two wells of the potential are also
    reported.}
  \begin{tabular*}{0.48\textwidth}{@{\extracolsep{\fill}} cccccc}
    \hline
 $\alpha_0$ &  $\alpha_1$ &  $\alpha_2$  & $\omega_{01}$ &$N_\ell$& $N_r$ \\
  (au) &  (au) & (au) & (cm$^{-1}$) &   &  \\
 \hline
 1/2048 & 1/256 & 2~10$^{-3}$ &  557 & 6 & 2\\
 \hline
 \hline
  $\gamma_1$  & $\delta_1$  & $\gamma_2$ &$\delta_2$  & & \\
  (au) & (au) &   (au) &  (au) &  &\\
  \hline
 1 & 2 & 1 & 2 &  & \\
   \hline
  \end{tabular*}
   {\label{tab:table1}}
\end{table}

\begin{figure}[t]
  \centerline{\includegraphics[width=8.3cm]{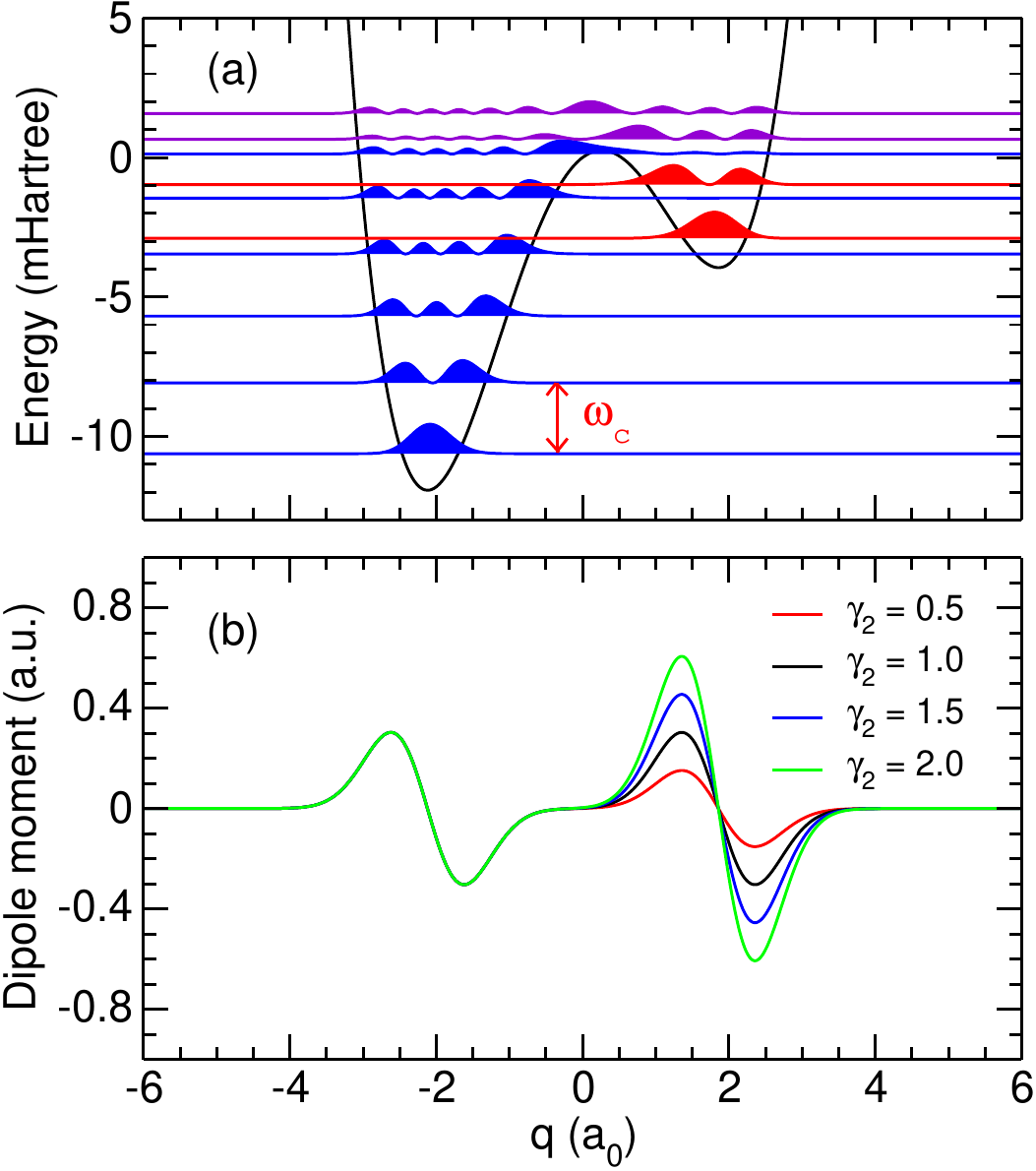}}
  \caption{ \label{Fig:Potential} (a) Potential energy curve and
    probability densities of the ten lowest molecular eigenstates
    obtained using the parameters given in
    Table~\ref{tab:table1}. Probability densities located in the left
    or right well are represented in blue and red, respectively, while
    the densities spreading over the barrier are represented in
    purple;
    (b) Dipole moment functions obtained using the parameters given in
    Table~\ref{tab:table1} and {several values of $\gamma_2$ given in
    atomic units.}}
\end{figure}

The vibrational Hamiltonian assumes the following asymmetric
double-well potential mimicking a bistable molecule
\begin{equation}
\tilde{V}(q) = \alpha_0 \, q^4 - \alpha_1 \, q^2 \, + \alpha_2 \, q ,
\label{Eq:Pot}
\end{equation}
in which $\alpha_0>0$, $\alpha_1>0$ and $\alpha_2$ are three
parameters. The positions of the two local minima of this double-well
potential are denoted as $q_1$ and $q_2>q_1$. In the case of a
{symmetric} potential ($\alpha_2$=0), these equilibrium positions are
found at $q_2=-q_1=\sqrt{ {\alpha_1}/{2 \alpha_0}}$.
{As we are mainly interested in the phenomenology of
  polaritonic chemistry, we have chosen to use values for
  these quantities that do not relate to any particular chemical system, but
  yield molecular positions, transition frequencies, and depths of the potential wells that are of the same orders of magnitude as realistic molecules, namely Angstr\"oms, hundreds of cm$^{-1}$, and fractions of eVs. These values, reported in Table \ref{tab:table1}, yield the potential energy curve represented in Fig.~\ref{Fig:Potential}(a).}
The discrete variable representation (DVR)\cite{Lill1982,Lill1985} was
used to solve the 1D time-independent Schr\"odinger equation and
determine the first bound states of the double-well asymmetric
potential, {assuming the same reduced mass of $\mu=1.43$~amu as employed by Triana {\em et al.}\cite{Triana2020}  With the current parameters, the vibrational frequency $\omega_{01}$ of the deepest well is found at 557~cm$^{-1}$, the second well having a lower fundamental frequency of 422.5~cm$^{-1}$. The bound states of the asymmetric potential are also depicted in Fig.~\ref{Fig:Potential}(a) with different colors for the two wells.}
Following this representation, the two wells will be referred to as
left ($\ell$) and right ($r$). The vibrational states corresponding to
the left (resp. right) well will be denoted as $v_{\ell}$
(resp. $v_r$).  With the current parameters, the left well is the
deepest one and supports $N_\ell= 6$ vibrational states, while the
right well contains $N_r = 2$ states. Higher vibrational states spread
over the barrier and are not confined to either well. The cavity mode
being in resonance with the first allowed vibrational transition, it
corresponds to the transition between the two first levels of the left
well, {\it i.e.} {$ \omega_c=  \omega_{01} =$
557~cm$^{-1}$}.
%

\subsection{Dipole moment function}

To describe the light-matter interaction, {we assume
  for simplicity that the molecule is non-polar in both wells and
  extend the work of Hernandez and Herrera\cite{Hernandez2019} to a double well-potential to obtain the empirical dipole moment function} 
\begin{equation}
\tilde{d}(q) = -\gamma_1 \, (q-q_1) \, e^{-\delta_1 (q-q_1)^2} -
\gamma_2 \, (q-q_2) \, e^{-\delta_2 (q-q_2)^2},
\label{Eq:dipole}
\end{equation}
with $\gamma_k$ and $\delta_k$ some (strictly positive) parameters for
$k=1,2$. This functional form is appropriate for a molecular system
that is nonpolar in the vicinity of both potential wells ($d(q_k)= 0$
for $k=1,2$) {and it is also well behaved with respect to a number of physically expected rules such as being maximum at a finite coordinate around equilibrium and vanishing at long distance.}\cite{Hernandez2019} The polarizability $\gamma_k$ measuring the dipole gradient at the corresponding minimum $q_k$ and governing the dipole
amplitude the associated well being an adjustable parameter. Using the
values reported in Table \ref{tab:table1}, the variations of the
dipole moment function with coordinate $q$ are shown in
Fig.~\ref{Fig:Potential}(b). In the following, the reference dipole
moment function will be the one defined by $\gamma_2 = \gamma_1 =$ {1.0~au.} Other dipole moment functions obtained with
alternative values of $\gamma_2$ are displayed in
Fig.~\ref{Fig:Potential}(b). As the ratio $\gamma_2/\gamma_1$ governs
the relative excitation probabilities in the left and the right wells,
the sensitivity of our results with respect to this ratio will later
be investigated.

\subsection{Polaritonic state properties}

The hybrid light-matter polaritonic states $ |P_i\rangle$ are obtained by diagonalization of the total Hamiltonian $\tilde{H}$  using a finite number of {vibrational states $N_v$ and photonic states $N_p$.} These states can be written in the uncoupled basis set $|v,n \rangle$ as
\begin{equation}
 |  P_i\rangle = \sum_{v = 0}^{N_v-1} \sum_{n = 0}^{N_p-1}  c_{v,n}^{(i)} \,  |  v,n \rangle.
\label{Eq:coeff}
\end{equation}
The associated eigenenergies will be denoted as $\varepsilon_i$. The vibrational polaritonic states themselves are more conveniently characterized by appropriate averaging over light or matter degrees of freedom. For each polaritonic state, $  |  P_i\rangle$, the mean value of $q$, noted $\langle q \rangle_i$ 
and the average number of photons $\langle n_p\rangle_i$ are computed as 
\begin{align}
\langle q \rangle_i &= \langle P_i  | q  |  P_i \rangle, \\
\langle n_p\rangle_i &= \langle P_i  |  \tilde{a}^\dagger \tilde{a}  |  P_i \rangle.
\end{align}
%
The assignment of a given polaritonic state  $|P_i\rangle$ within one of the two wells ($\ell$ or $r$) is obtained from the value of $\langle q\rangle_i$. For a symmetric well the top of the barrier separating the two wells is at $q = 0$, therefore the left well is characterized by $\langle q\rangle < 0$ and the right one by $\langle q\rangle > 0$. This remains essentially true for the asymmetric well considered here as the top of the barrier is only shifted by about {0.26~bohr}. The sign of $\langle q\rangle_i$ will thus be used to determine the localization of the associated polariton $\ket{P_i}$.

\subsection{Infrared spectroscopy}
The strength of the radiative coupling between two polaritonic states $ |  P_i\rangle$ and $ |  P_j\rangle$ can be characterized by
\begin{equation}
R_{i,j} =  \left|  \frac{\langle P_i  |  \tilde{d}  |   P_j\rangle}{\langle 0_\ell  |  \tilde{d}  | 1_\ell \rangle}  \right| ^2,
\end{equation}
where, as in Eq.~(\ref{Eq:H_vn}), we choose to normalize the coupling between the two polaritons by the corresponding radiative coupling of the first IR transition. 

At thermal equilibrium, the state of the confined system can be probed using IR spectroscopy. Denoting by $\beta=1/k_{\rm B}T$ the inverse temperature (with $k_{\rm B}$ the Boltzmann constant),  the linear absorption intensity obtained with an excitation laser at frequency $\omega$ and characterized by a Gaussian spectral profile can be estimated by
\begin{equation}
I_\beta(\omega) \propto    \sum_i \sum_{j>i} \frac{ (e^{-\beta \epsilon_i} - e^{-\beta\epsilon_j}) }{Z(\beta)} \, \,  R_{i,j}  \times g(\omega-\omega_{ij}) 
\label{eq:IR_spectra_temp}
\end{equation}
with $ \omega_{ij}=\varepsilon_j-\varepsilon_i$, $Z(\beta)$ the canonical partition function and $g(\omega-\omega_{ij})$ the Gaussian broadening function centered at $\omega_{ij}$. 
The IR absorption spectrum depends on the coupling parameter $\lambda_c$ through the eigenenergies $\epsilon_i$ and the radiative couplings $R_{i,j}$. The spectral response of the hybrid system will be given by Eq.~(\ref{eq:IR_spectra_temp}) only if 
the polaritonic state relaxation time is longer than the temporal profile of the laser excitation.

\subsection{Polaritonic state evolution}

The time evolution of an initial non-stationary state, denoted as $|\Psi(t=0)\rangle$, can be obtained by solving the time-dependent Schr\"odinger equation (TDSE) at a given value of the coupling parameter $\lambda_c$. From a unitary transformation, the initial wavepacket can be decomposed linearly over the polaritonic states as


 
\begin{equation}
 |  \Psi (t=0)\rangle =   \sum_i  b_i(0)  \  |  P_i\rangle,
 \label{Eq:initial}
\end{equation}
where the coefficients $b_i(0)$ depend on $\lambda_c$.
The time evolution of the wavepacket is then governed by the TDSE, which can be solved exactly in the polaritonic basis set as the Hamiltonian does not explicitly depend on time.
The wavepacket at time $t$ is more conveniently expressed in the uncoupled basis set as
\begin{equation}
 |  \Psi (t)\rangle =   \sum_{v=0}^{N_v-1} \sum_{n=0}^{N_p-1}  \underbrace{ \sum_k b_k(0) \, e^{-i\epsilon_kt} \, c_{v,n}^{(k)} }_{A_{v,n}(t)}    \,   | v, n\rangle,
\end{equation}
where $  |  {A_{v,n}(t)}  | ^2$ corresponds to the probability for the system to lie in the uncoupled state $ |v,n\rangle$ at time $t$.
The mean values $\langle q \rangle(t) = \langle \Psi (t)  |  q  |  \Psi (t)\rangle$ and $\langle n_p \rangle(t) = \langle \Psi (t)  |  \tilde{a}^\dagger \tilde{a}  |  \Psi (t)\rangle$ can be computed along the trajectory to analyze the position of the system inside the potential wells and the number of photons in the cavity, respectively.

\section{Results and discussion}
\label{sec:Results}

All the results presented below were obtained {for a single
  molecule ($N=1$) with maximum numbers of $N_v=40$ vibrational states
  and $N_p=40$ photons. These values ensure} convergence for the
polaritonic states studied here. Unless otherwise
specified, the parameters used for the potential function are
{those reported} in Table~\ref{tab:table1}, notably $\gamma_2 =
\gamma_1=${1.0~au}.

\subsection{Avoided crossings and polariton localization}
\begin{figure}[t]
  \centerline{\includegraphics[width=8.3cm]{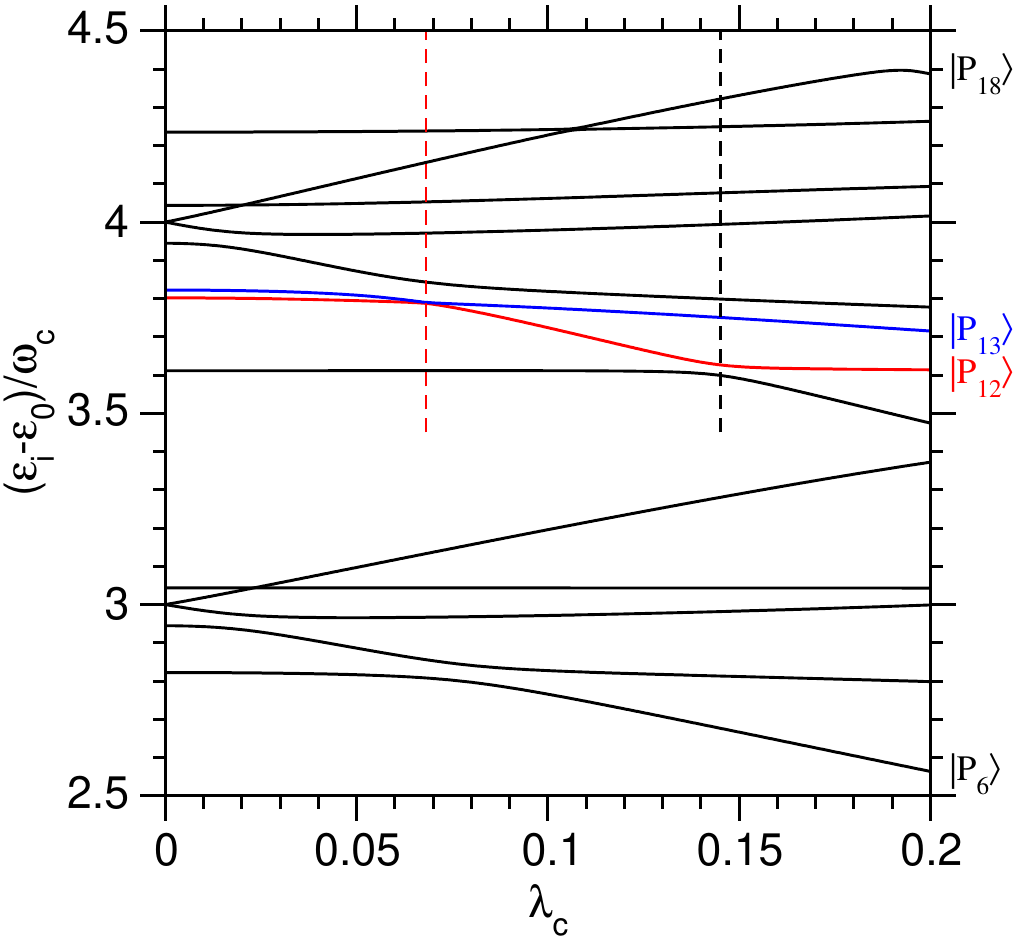}}
  \caption{Eigenenergies of the polaritonic states $|P_i\rangle$,
    $i=6$--18, lying in the region near $3\omega_c$ and $4\omega_c$
    above the ground polaritonic state, as a function of the coupling
    strength $\lambda_c$. The eigenenergies of states $|P_{12}\rangle$
    and $|P_{13}\rangle$ are highlighted in red and blue,
    respectively. The dashed vertical lines highlight the two avoided
    crossings involving the pairs $(|P_{11}\rangle,|P_{12}\rangle)$
    and $(|P_{12}\rangle,|P_{13}\rangle)$ and occurring at
    $\lambda_c=0.145$ and 0.068, respectively.}
  \label{Fig:eigenvalue} 
\end{figure}

The role of the coupling parameter $\lambda_c$ on vibrational
polaritonic states is first discussed, with Fig.~\ref{Fig:eigenvalue}
showing the variations of their eigenenergies (relative to the ground
polaritonic state) with increasing values of $\lambda_c$. At each
value of the coupling parameter, the polaritonic states $\ket{P_i}$
are ordered by increasing energy. Since the investigated system
features some true crossings, the label $i$ associated with a given
physical state may change depending on the value of $\lambda_c$.
\begin{table}[b]
\small
  \caption{Assignement of the polaritonic states ($| P_i\rangle$, $i=
    6$--18) for $\lambda_c$=0. The symbols $\ell$ and $r$ correspond
    to vibrational states localized in the left and right potential
    wells, respectively}
  \begin{tabular*}{0.48\textwidth}{@{\extracolsep{\fill}}ccc}
    \hline
Polariton & $\varepsilon_i-\varepsilon_0 \sim 3\omega_c$ & $\varepsilon_i-\varepsilon_0 \sim 4\omega_c$\\
$ |  P_i\rangle$ &  $ |  v,n\rangle$ & $ |  v,n\rangle$\\
\hline
$ |  P_6\rangle$ & $   | 3_\ell,0\rangle$  & \\
$ |  P_7\rangle$ &  $    |  2_\ell,1\rangle$ & \\
$ |  P_8\rangle$ &  $   | 1_\ell,2\rangle$ & \\
$ |  P_9\rangle$ &     $ |  0_\ell,3\rangle$ & \\
$ |  P_{10}\rangle$ & $  |  0_r,0\rangle$ & \\
\hline 
$ |  P_{11}\rangle$ &  & $   |  4_\ell,0\rangle$\\
$ |  P_{12}\rangle$ &  & $   | 1_r,0\rangle$\\
$ |  P_{13}\rangle$ & & $   | 3_\ell,1\rangle$      \\
$ | P_{14}\rangle$ &  &   $   |  2_\ell,2\rangle$  \\
$ |  P_{15}\rangle$ &  & $  | 1_\ell,3\rangle$   \\
$ |  P_{16}\rangle$ &  & $   | 0_\ell,4\rangle$ \\
$ |  P_{17}\rangle$ &  &     $  |  0_r,1\rangle$\\
$ |  P_{18}\rangle$ &  &  $  | 5_\ell,0\rangle$    \\
    \hline
  \end{tabular*}
   {\label{tab:table2}}
\end{table}
To study coherent switching between the two wells, we focus on the
energy range where both potential wells exist, while remaining below
the barrier.  As seen from Fig.~\ref{Fig:Potential}(a), this typically
corresponds to energies between $3 \omega_c$ and $4 \omega_c$ above
the ground state.  As shown by the corresponding assignments in
Table~\ref{tab:table2}, at $\lambda_c = 0$ five polaritonic states
($|P_i\rangle$ for $i=6$--10) are located near $3 \omega_c$, while
eight states ($i=11$--18) reside near $4 \omega_c$.
%
%
In this range, the eigenenergies of the polaritonic states are found to be generally sensitive to $\lambda_c$, with neighboring states often exhibiting avoided crossings.

As seen in Fig.~\ref{Fig:eigenvalue}, several polaritonic states are
doubly degenerated at $\lambda_c=0$, as the consequence of the exact
resonance condition between $\omega_c$ and the first vibrational transition of the left potential well being met. As an example,
$|P_8\rangle$ and $|P_9\rangle$ share the same energy at $\lambda_c=0$
as they correspond to $|0_\ell,3\rangle$ and $|1_\ell,2\rangle$,
respectively [see Table~\ref{tab:table2}]. Note that the energies of
states $|2_\ell,1\rangle$ and $|3_\ell,0\rangle$ are shifted from the
doubly degenerated value due to the anharmonicity of the potential
well, which causes a detuning between the cavity mode frequency and
the vibrational hot bands.  A similar degeneracy occurs between $|
P_{15}\rangle$ and $| P_{16}\rangle$, which respectively correspond to
$|0_\ell,4\rangle$ and $|1_\ell,3\rangle$ at $\lambda_c=0$.

Several avoided crossings are also visible in Fig.~\ref{Fig:eigenvalue}. As discussed in Ref.~\citenum{Hernandez2019}, they occur when the main contributions in the $ | v,n\rangle$ basis set of the two polaritonic states involved are characterized by $ | \Delta n | =1$. Near $3 \omega_c$, avoided crossings involve the ($ | P_6\rangle$, $ | P_7\rangle$) pair
at $\lambda_c \simeq 0.08$, as well as the ($| P_7\rangle$, $|P_8\rangle$) pair at $\lambda_c \simeq 0.02$. 
The decomposition of these three polaritonic states in the uncoupled basis set shows that they only involve vibrational states sitting in the left potential well ($\langle q \rangle_i < 0$ for all values of $\lambda_c$), making those avoided crossings unsuitable to achieve coherent switching between the two wells.

\begin{figure}[t]
  \centerline{\includegraphics[width=8.3cm]{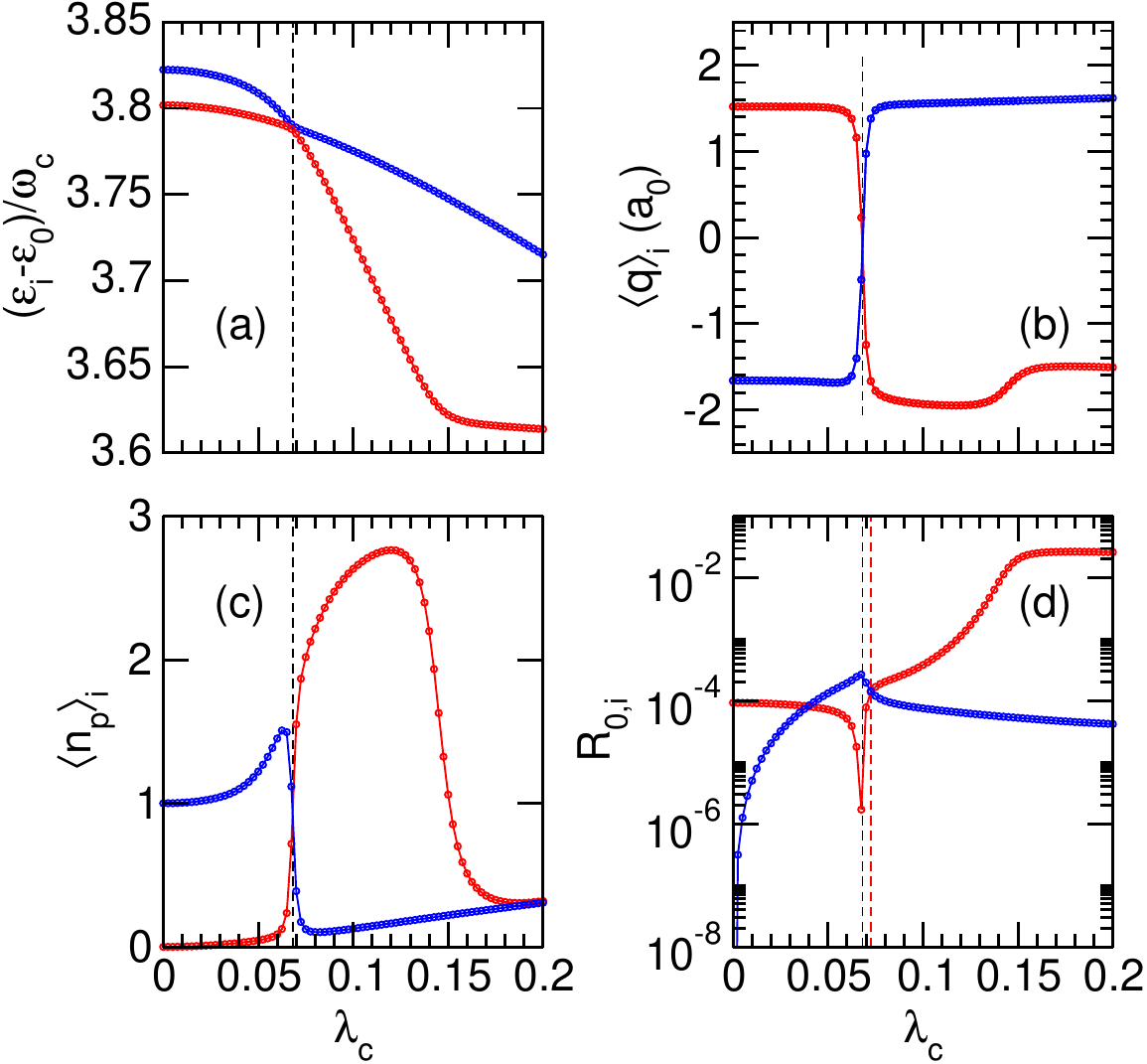}}
 \caption{Variations of stationary properties of the polaritonic states {$|P_i \rangle$ for $i=12$ (red) and $i=13$ (blue)} 
 with increasing coupling strength $\lambda_c$. (a) Relative energies $(\varepsilon_i - \varepsilon_0)/\omega_c$; (b) Average position $\langle q \rangle_i$ along the potential wells; (c) Average number of photons $\langle n_p \rangle_i$; (d) Normalized
   radiative coupling strength $R_{0,i}$. In the four panels, the black vertical dashed lines highlight the position of the avoided crossing at
    $\lambda_c=\lambda_{\rm crit}$. In panel (d), the red vertical dashed line at $\lambda_c=\lambda_R$ highlights the coupling strength at which $R_{0,12}=R_{0,13}$.}
 \label{Fig:P12_P13}
\end{figure}
Two other avoided crossings can be found around $4 \omega_c$: between $ | P_{12}\rangle$ and $ |P_{13}\rangle$ at $\lambda_c=0.068$ and between $ | P_{11}\rangle$ and $ |P_{12}\rangle$ at $\lambda_c=0.145$. Fig.~\ref{Fig:P12_P13} focuses on the ($|P_{12}\rangle$, $|P_{13}\rangle$) pair and shows the variations of several properties associated with these two polaritonic states when $\lambda_c$ increases, starting with their relative energies in Fig.~\ref{Fig:P12_P13}(a).  
The corresponding variations of the average position $\langle q\rangle_i$ and photon number $\langle n_p\rangle_i$  are represented in panels (b) and (c) of the same figure.
The first avoided crossing occurs at $\lambda_{\rm crit}=0.068$ where the energy difference between $ | P_{12}\rangle$ and $ |P_{13}\rangle$ is as small as $\Delta \varepsilon_{\rm crit}/\omega_c \approx 3\times 10^{-3}$.
As seen in Fig.~\ref{Fig:P12_P13}(b), the system effectively switches between the two wells near this avoided crossing, with simultaneous switching of $\ket{P_{12}}$ from the right well ($\langle q\rangle_{12} > 0$ for $\lambda_c < \lambda_{\rm crit}$) to the left well ($\langle q\rangle_{12} < 0$ for $\lambda_c > \lambda_{\rm crit}$), and of $\ket{P_{13}}$ from left to right. 
The avoided crossing also strongly alters the average number of photons $\langle n_p \rangle$ in the two polaritonic states with $|P_{12}\rangle$ suddenly gaining two photons and $|P_{13}\rangle$ losing one photon.

The avoided crossing between $\ket{P_{11}}$ and $\ket{P_{12}}$ at $\lambda_c=0.145$ induces only a minor shift in the energy and average position of the $|P_{12}\rangle$ state [see Fig.~\ref{Fig:P12_P13}(a) and (b)]. In particular, the system does not switch between the two wells at this value of the coupling strength {($\langle q\rangle_{12}$ does not change sign) as both $\ket{P_{11}}$ and $\ket{P_{12}}$ reside in the left well when approaching $\lambda_c=0.145$.} 
Only the photon number varies significantly near $\lambda_c=0.145$, thus suggesting that this avoided crossing is due to a photon exchange between $\ket{P_{11}}$ and $\ket{P_{12}}$ rather than a coherent switching between the two wells. 

%
\begin{figure}[t]
  \centerline{\includegraphics[width=8.3cm]{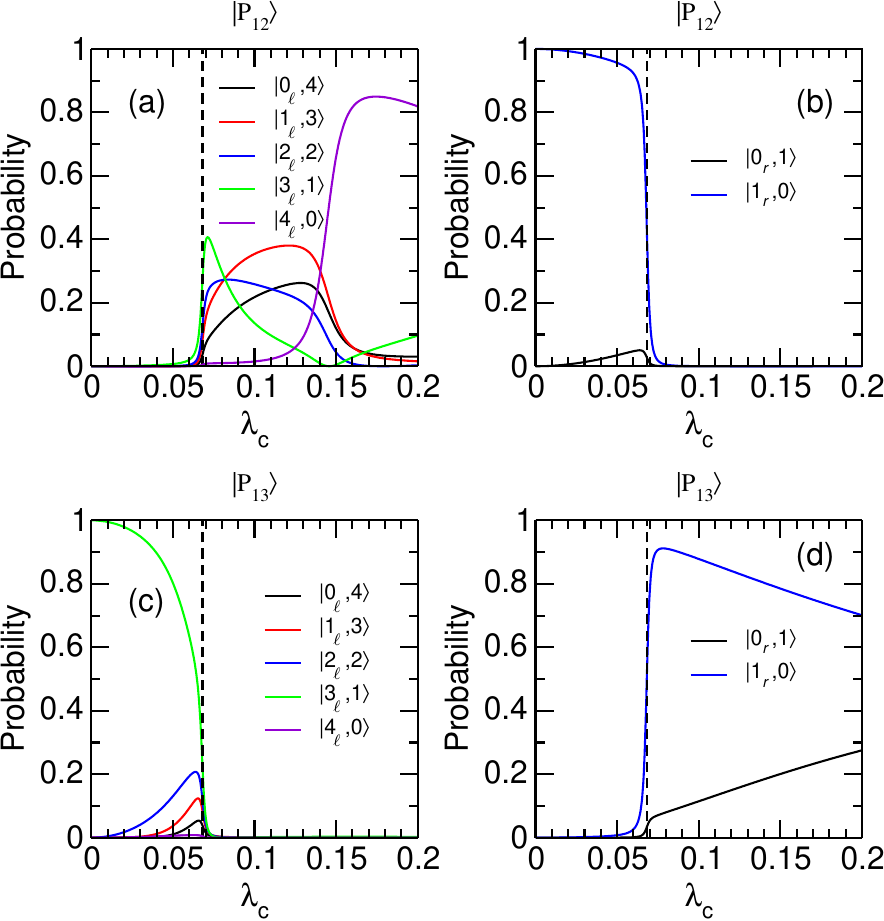}}
  \caption{Probabilities for polaritonic states $|P_i\rangle$, $i=12$
    and 13 to be in the uncoupled states $|v_{\ell/r},n\rangle$, as a function of the coupling strength $\lambda_c$. (a) and (b) $|\langle v,n | P_{12}
    \rangle|^2$; (c) and (d) $|\langle v,n | P_{13}
    \rangle|^2$. Uncoupled states that belong to the left (right) well
    are shown in the left (right) panels. The vertical dashed lines
    highlight the position of the avoided crossing at
    $\lambda_c=\lambda_{\rm crit}$.} 
  \label{Fig:Sup2} 
\end{figure}
To better understand the evolution of the system 
near the avoided crossing at
$\lambda_{\rm crit} = 0.068$, which is the only one leading to a switching between the two wells, we consider the decomposition
of polaritonic states $ | P_{12}\rangle$ and $ |P_{13}\rangle$ in the $ | v,n \rangle$ basis set. The weights of $ |P_{12}\rangle$ and $|P_{13}\rangle$ on different uncoupled states are given in Fig.~\ref{Fig:Sup2} as a function of $\lambda_c$.
The distribution of the polaritonic states
among the two potential wells is found to vary abruptly 
near $\lambda_{\rm crit}$, highlighting the
extreme sensitivity of the coherent switching process to the strength of the light-matter coupling. As
$\lambda_c$ crosses $\lambda_{\rm crit}$ from below, the polaritonic state $|P_{12}\rangle$ which mostly involved the first excited state of the right well with no additional photon ($\ket{1_r,0}$), turns
into a mixture of vibrationally excited states of the
left well with $v_\ell=0$--4 and a complementary number $n_p =4-v_\ell$ of photons [see Fig.~\ref{Fig:Sup2}(a)-(b)]. The significant weights carried by states with $n_p >1$ explain the sudden increase in the average number of photons of $\ket{P_{12}}$ found in Fig.~\ref{Fig:P12_P13}(c). Concomitantly, the polaritonic state $|P_{13}\rangle$ that occupied the left well in its third vibrationally excited state with one photon in the cavity ($\ket{3_\ell,1}$), is entirely transferred to the right well as a superposition of states $\ket{1_r,0}$ and $\ket{0_r,1}$ [see Fig.~\ref{Fig:Sup2}(c)-(d)]. This superposition is first dominated by the $\ket{1_r,0}$ state while state $\ket{0_r,1}$ gains in weight as $\lambda_c$ increases. This trend is consistent with the evolution of the average number of photons in $\ket{P_{13}}$ reported in Fig.~\ref{Fig:P12_P13}(c).

Considering the strong sensitivity of this phenomenon with respect to the coupling strength, the influence of the dipole moment function was also addressed by changing the value of $\gamma_2$ in Eq.~(\ref{Eq:dipole}), leaving the other parameters unaffected. 
The same phenomenology was obtained for all the studied values of $\gamma_2$, and the critical coupling strengths $\lambda_{\rm crit}$ 
computed for $\gamma_2$ varying in the range {0.5--2.0~au} are given in Table~\ref{tab:table3}.
\begin{table}[b]
\small
  \caption{Critical coupling strengths $\lambda_{\rm crit}$ and
    $\lambda_R$ involving the polaritonic states
    $|P_{12}\rangle$ and $|P_{13}\rangle$, as computed for increasing
    values of $\gamma_2$. The energy gaps $\Delta
    \varepsilon_{\rm crit}/\omega_c$ and $\Delta \varepsilon_R/\omega_c$ refer to the energy difference
    between these two polaritonic states when $\lambda_c = \lambda_{\rm
      crit}$ and $\lambda_c=\lambda_R$, respectively}
  \begin{tabular*}{0.48\textwidth}{@{\extracolsep{\fill}}ccccc}
    \hline
 $\gamma_2$ &  $\lambda_{\rm crit}$ & $\Delta \varepsilon_{\rm crit}/\omega_c$ & $\lambda_R$ & $\Delta \varepsilon_R/\omega_c$ \\
  (au) & ($\times 10^{-3}$)  & ($\times10^{-3}$) & ($\times10^{-3}$) & ($\times10^{-3}$) \\
    \hline
  0.50  & 61.9 & 2.8 & 67.8 & 7.7 \\
  0.75  & 64.4 & 2.8 & 69.7 & 7.0 \\
  1.00  & 68.1 & 2.7 & 72.5 & 6.2 \\
  1.25  & 72.5 & 2.6 & 76.3 & 5.3 \\
  1.50  & 78.1 & 2.4 & 81.3 & 4.6 \\
  1.75  & 85.0 & 2.2 & 87.2 & 3.7 \\
  2.00  & 92.8 & 2.1 & 94.4 & 3.0 \\
    \hline
  \end{tabular*}
   {\label{tab:table3} }
\end{table}
Since $\gamma_2$ governs the dipole moment amplitude in the right well, the states $|  0_r,1 \rangle$ and $|1_r,0\rangle$ become increasingly coupled as this parameter increases.
For small values of the coupling strength ($\lambda_{c} <\lambda_{\rm crit}$), the state $|1_r,0\rangle$ is the main component of polariton $|P_{12}\rangle$ [see Fig.~\ref{Fig:Sup2}(b)] and $|0_r,1 \rangle$ is a major component of $|P_{17}\rangle$. The increasing coupling between these two states leads them to repel one another, and since $|P_{17}\rangle$ has a higher energy than $|P_{12}\rangle$, this stabilizes $|P_{12}\rangle$ by lowering its energy. In contrast, the polaritonic state $|P_{13}\rangle$ remains mostly unaffected as it lies in the left well. As seen in Table~\ref{tab:table3}, this leads to an increase of the critical value $\lambda_{\rm crit}$ of the avoided crossing upon increasing $\gamma_2$. We also find that the energy difference $\Delta \varepsilon_{\rm crit}$ at the critical coupling strength decreases with increasing values of $\gamma_2$. Above the avoided crossing ($\lambda_{c} >\lambda_{\rm crit}$) $|P_{12}\rangle$  and $|P_{13}\rangle$ switch between wells, hence for larger coupling strengths $|P_{12}\rangle$ is no longer affected by the value of $\gamma_2$ while $|P_{13}\rangle$ is stabilized when this parameter increases.

Finally, the radiative coupling $R_{0,i}$ between the polaritonic ground state $\ket{P_0}$ and a given $\ket{P_i}$ ($i = 12$, 13) state is displayed in Fig.~\ref{Fig:P12_P13}(d) as a function of increasing coupling strength $\lambda_c$. 
In the limit of small values of $\lambda_c$, the radiative coupling
ratios $R_{0,12}$ and $R_{0,13}$ converge toward the values in the
free molecule, namely $|\langle 0_\ell|\tilde{d}| 1_r\rangle |^2 /
|\langle 0_\ell|\tilde{d}|1_\ell\rangle|^2\simeq 10^{-4}$ for the
$|P_{12}\rangle$ state and $\langle 0_r,0 | \tilde{d} | 3_\ell,1\rangle = \langle 0 _r| \tilde{d} | 3_\ell\rangle \times \langle 0
| 1\rangle =0$ for $|P_{13}\rangle$.
At higher values of $\lambda_c$, and especially near the critical coupling strength $\lambda_{\rm crit}$, the variations of the radiative coupling are quite strong [notice the logarithmic scale in Fig.~\ref{Fig:P12_P13}(d)]. Another interesting feature is found for $\lambda_c=0.0725=\lambda_R$ at which the two states $|P_{12}\rangle$ and $|P_{13}\rangle$ share the same radiative coupling strength. This particular value of $\lambda_c$ is important since it allows the system to be prepared as a coherent superposition of these two polaritonic states, thus allowing some manipulation with a laser field. This possibility will be explored further and discussed in section~\ref{sssec:laser_pulse}.
The values of $\lambda_R$ and the corresponding energy gaps $\Delta \varepsilon_R$ are also reported in Table~\ref{tab:table3} as a function of increasing $\gamma_2$. As is the case with $\lambda_{\rm crit}$, $\lambda_R$ increases with $\gamma_2$, owing to greater resulting coupling in the right well. The associated energy gap $\Delta \varepsilon_R$ is found to decrease when $\gamma_2$ increases, which we speculatively interpret as the consequence of the avoided crossing becoming increasingly sharper with increasing transition dipole moment.

\subsection{Coherent state switching}


\subsubsection{Starting from a free excited molecule}

\begin{figure}[t]
  \centerline{\includegraphics[width=8.3cm]{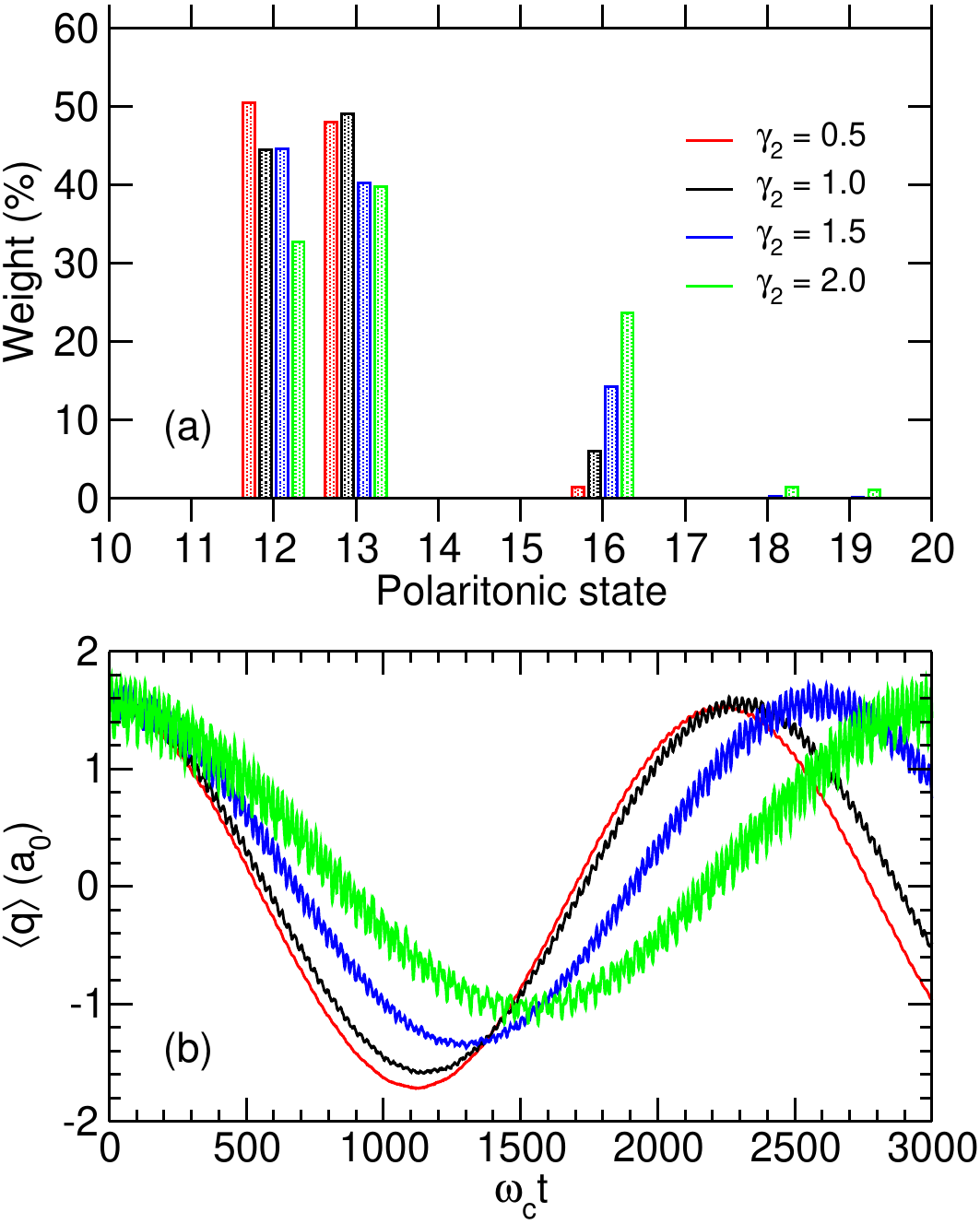}}
  \caption{ (a) Initial weights $|b_i(0)|^2$ of the
    polaritonic states $|P_i\rangle$ for $i=10$--20 and for
    four values of $\gamma_2$ given in {atomics units}; (b) Average position $\langle q\rangle$ as a function of
    time for $\lambda_c$ = $\lambda_{\rm crit}$, assuming the
    system is initiated at $t=0$ in the state $| 1_r,0\rangle$, for the same values of $\gamma_2$.}
  \label{Fig:v1-right_time} 
\end{figure}
We first consider a molecule prepared in a specific vibrational excited state $\ket{ v = 1_r}$ and that enters the {initially photon-less cavity} ($n_p = 0$) at time $t=0$, leading to the initial state $| 1_r,0\rangle$.
The coupling strength is taken at the critical value $\lambda_c = \lambda_{\rm crit}$ of the avoided crossing discussed above. At this value of the coupling strength the two polaritonic states $|P_{12}\rangle$ and $|P_{13}\rangle$ involved in the possible switching have a nonzero contribution on the initial state $| 1_r,0\rangle$ state [see
Fig.~\ref{Fig:Sup2}]. 
As seen in Fig.~\ref{Fig:v1-right_time}(a) for several dipole function parameters $\gamma_2$, the initial state is almost a linear combination of $ | P_{12} \rangle$ and $ | P_{13} \rangle$. This is especially true for small values of $\gamma_2$ where the mixing with other polaritonic states residing in the right well is minimal.

The time evolution of the wavepacket {in the initially photon-less cavity} is then followed by numerically solving the TDSE, starting from $ |1_r,0\rangle$.
The effects of inserting the molecule in the microcavity at $\lambda_c = \lambda_{\rm crit}$  are particularly spectacular with the system alternatively switching between the two wells with a period ranging from {around 2200 to 3000 $\omega_c t$, which using the parameters of Table~\ref{tab:table1} corresponds to 20--30~ps,} depending on the value of $\gamma_2$. This can be seen from the variations of $\langle q\rangle(t)$ displayed in Fig.~\ref{Fig:v1-right_time}(b). Note that when varying $\gamma_2$, the values of the coupling strength were adjusted to the corresponding critical values $\lambda_{\rm crit}$ as given in Table~\ref{tab:table3}. 

By increasing $\gamma_2$, the switching becomes less efficient and is associated with a longer period, as well as more significant fluctuations that
are a manifestation of the importance of polaritonic states other
than $|P_{12}\rangle$ and $|P_{13}\rangle$ in the decomposition of the initial state. 
The histogram of Fig.~\ref{Fig:v1-right_time}(a) reveals that those additional contributions are mostly due to state $\ket{P_{16}}$, which plays an important role when $\gamma_2$ increases and thus contributes to the additional high-frequency oscillations seen in Fig.~\ref{Fig:v1-right_time}(b). At $\lambda_c = 0$, $\ket{P_{16}}$ is referred to as $\ket{P_{17}}$ due to the true crossing occurring between $\lambda_c = 0$ and $\lambda_{\rm crit}$ and corresponds to the uncoupled state $\ket{0_r,1}$. However, it progressively gains weight on state $\ket{1_r,0}$ as $\lambda_c$ increases. The coupling between these two states being stronger when the dipole moment in the right well increases, the weight of polaritonic state $\ket{P_{16}}$ in the decomposition of {the initial state} $\ket{1_r,0}$ increases with $\gamma_2$.

The situation considered here of the specific preparation of the
system as it enters the microcavity is experimentally challenging, as
it would typically require the molecules to be prepared in a uniformly
excited flow perpendicular to the cavity axis. 
However, such experiments could be used
to prepare molecules in either well of a bistable potential by
adjusting their interaction time with the confined light. This could
be done, \textit{e.g.}, by tuning the velocity of the molecular flow
in order to synchronize the interaction time with the polaritonic
oscillation period, and might enable a higher degree of control of the
system state at the output of the cavity.
{As far as photon lifetimes are concerned, the switching period found with this model system would be far too long for liquid phase experiments, but is compatible with measurements in the gas phase.\cite{Wright2023}}
\subsubsection{Using a short IR laser pulse}
\label{sssec:laser_pulse}

As an alternative to the preparation of a specific vibrational state
outside of the cavity, we now consider the possibility of acting on a
system at thermal equilibrium inside the FP cavity through an external
excitation caused by a short IR laser pulse.
From Fig.~\ref{Fig:P12_P13}(d), the radiative couplings $R_{0,12}$ and
$R_{0,13}$ are found to cross each other at a coupling strength of
$\lambda_R=0.0725$ slightly higher than the critical value
$\lambda_{\rm crit}$ at which the corresponding eigenenergies
cross. At this higher coupling strength, the radiative excitation
probability from the ground polaritonic state towards $ | P_{12}
\rangle$ or $ | P_{13} \rangle$ become identical. This very specific
situation can be exploited to trigger dynamical coherent switching
between the polaritonic states $ | P_{12} \rangle$ and $ | P_{13}
\rangle$ by preparing the system in an equal superposition of these
two states using a short laser pulse: 
\begin{equation}
 | \Psi(t=0)\rangle \simeq \frac{1}{\sqrt{2}} \, ( | P_{12} \rangle +
 | P_{13} \rangle).
\label{Eq:WP}
\end{equation}

\begin{figure}[t]
  \centerline{\includegraphics[width=8.3cm]{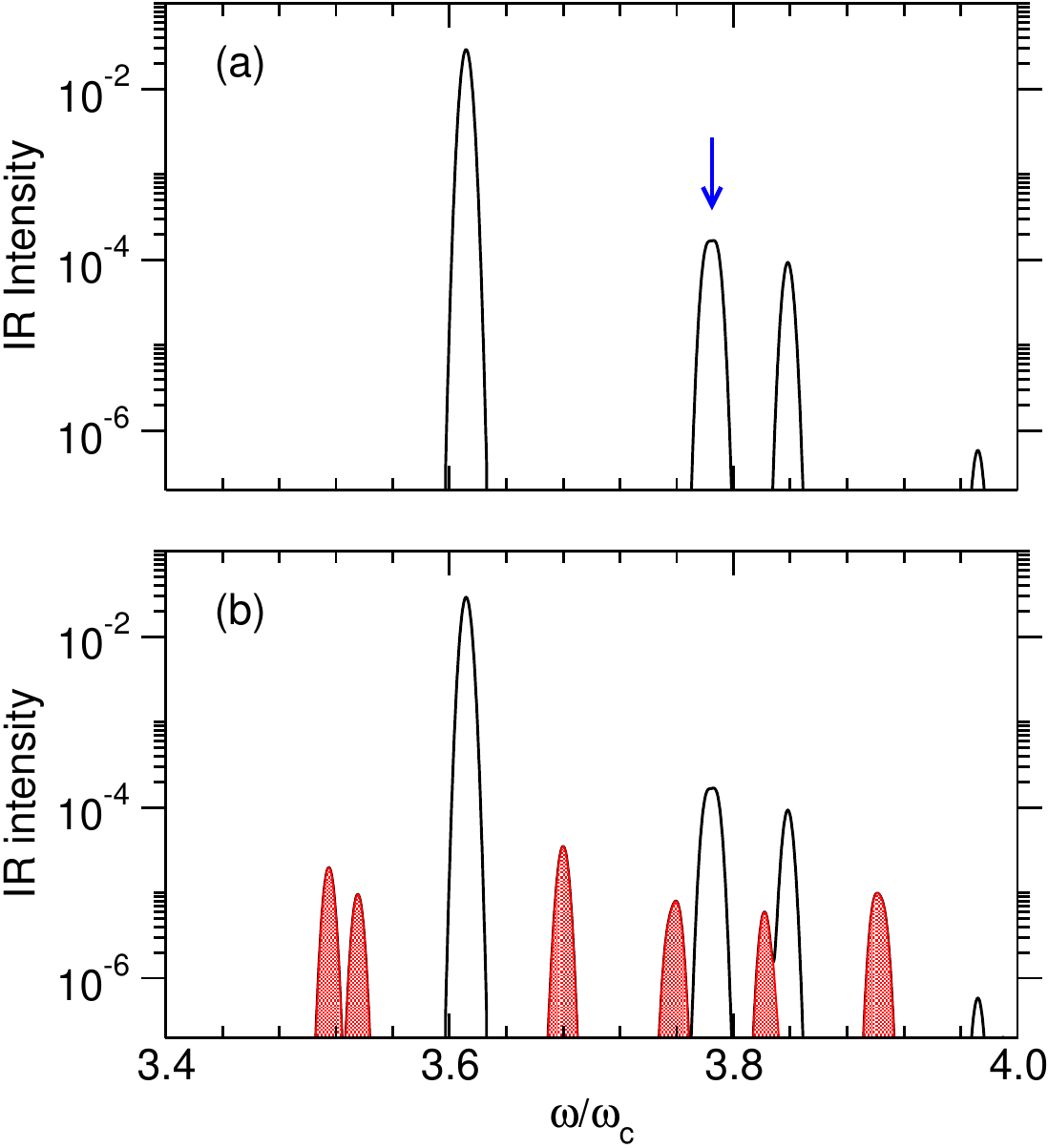}}
  \caption{Relative IR absorption intensity of the polaritonic system
    at thermal equilibrium obtained at the coupling strength
    $\lambda_c=\lambda_R$, and for
    {$\gamma_2=1.0$~au}. (a) $T=0.05\omega_c/k_{\rm
      B}$. The blue arrow highlights the excitation corresponding to
    the (almost identical) eigenenergies of the polaritonic states
    $|P_{12}\rangle$ and $|P_{13}\rangle$; (b) $T=0.15 \omega_c/k_{\rm
      B}$. The transitions corresponding to hot bands are shown in
    red. In both panels, the absorption spectra were convoluted with a
    Gaussian function having a width at half maximum of 4~cm$^{-1}$.}
 \label{Fig:Sup4}
\end{figure}
However, efficient switching requires strict control of this initial
state, which may be difficult to achieve at high temperatures.
Temperature effects were investigated specifically by calculating the
IR absorption spectra at two temperatures $T$ such that $k_B \,
T/\omega_c=0.05$ and 0.15, corresponding to about 40 and 120~K,
respectively. {Calculations were conducted with the set of parameters reported in Table~\ref{tab:table1}, including $\gamma_2=1.0$~au, and at the coupling strength $\lambda_c = \lambda_R$.} The relative IR absorption
intensities obtained from Eq.~(\ref{eq:IR_spectra_temp}) and
normalized by the intensity of the $v=0_\ell\to 1_\ell$ transition of
the free molecule are shown in Fig.~\ref{Fig:Sup4} at those two
temperatures.
At $T=0.05\omega_c/k_{\rm B}$, only transitions starting from the
ground polaritonic state can be distinguished. The most intense
transition is found at $\omega/\omega_c\approx 3.6$ and corresponds to
the overtone transition $v=0_\ell\to 4_\ell$ that connects the ground
polaritonic states $ |P_{0} \rangle$ to the excited state $| P_{11}
\rangle$. The two transitions leading from $ | P_{0} \rangle$ to
either $| P_{12}\rangle$ or $ | P_{13} \rangle$, which are nearly
degenerate at this value of the coupling strength, appear as a unique
feature around $\omega/\omega_c \approx 3.78$. At
$T=0.15\omega_c/k_{\rm B}$, extra lines appear
as the signature of hot bands. In particular, three transitions arise
at $\omega/\omega_c =3.75$ ($| P_{2} \rangle \rightarrow | P_{23}
\rangle$), 3.76 ($| P_{2} \rangle \rightarrow | P_{24} \rangle$) and
3.82 ($| P_{1} \rangle \rightarrow | P_{21} \rangle$), near the
targeted transition at $3.78\omega_c$ 
interest, and may interfere with the preparation of the desired
initial superposition, making it difficult to prepare the system at
this temperature. {Hence, for this model system $k_{\rm B}T/\omega_c=0.15$ appears as an approximate upper limit below which the polaritonic states $| P_{12}\rangle$ and $ | P_{13} \rangle$ can be superimposed in a coherent way, and above which other polaritonic states coming from hot bands would compromise this coherence.}

The IR spectra shown in Fig.~\ref{Fig:Sup4} are convoluted with a
Gaussian function having a width at half maximum of 4 cm$^{-1}$ {($\approx 7. 10^{-3}\omega_c)$} to
mimic the spectral width of a {15~ps} laser. In such
conditions the contributions of polaritonic states $|P_{12}\rangle$
and $|P_{13}\rangle$ cannot be distinguished and both states are
therefore excited by the laser pulse.  This is consistent with the
very small energy gap between these states at $\lambda_c=\lambda_R$
(with the present choice of parameters $\Delta \varepsilon_R\approx
3.5~$cm$^{-1}$). Therefore, at low temperatures and for an IR
excitation using a laser pulse with a spectral width of the order of
$\Delta \varepsilon_R$, the initial wavepacket $|\Psi(t=0)\rangle$ can
be prepared as $ | \Psi(t=0)\rangle \propto (\sqrt{R_{0,12}} \, |
P_{12} \rangle + \sqrt{R_{0,13}} \, | P_{13} \rangle$). Since
$R_{0,12}=R_{0,13}$ when $\lambda_c=\lambda_R$, the initial wavepacket
$|\Psi(t=0)\rangle$ can be rewritten as a coherent superposition,
%
$| \Psi(t=0)\rangle \simeq \frac{1}{\sqrt{2}} \, ( | P_{12} \rangle +
| P_{13} \rangle)$.
%

\begin{figure}[t]
 \centerline{\includegraphics[width=8.3cm]{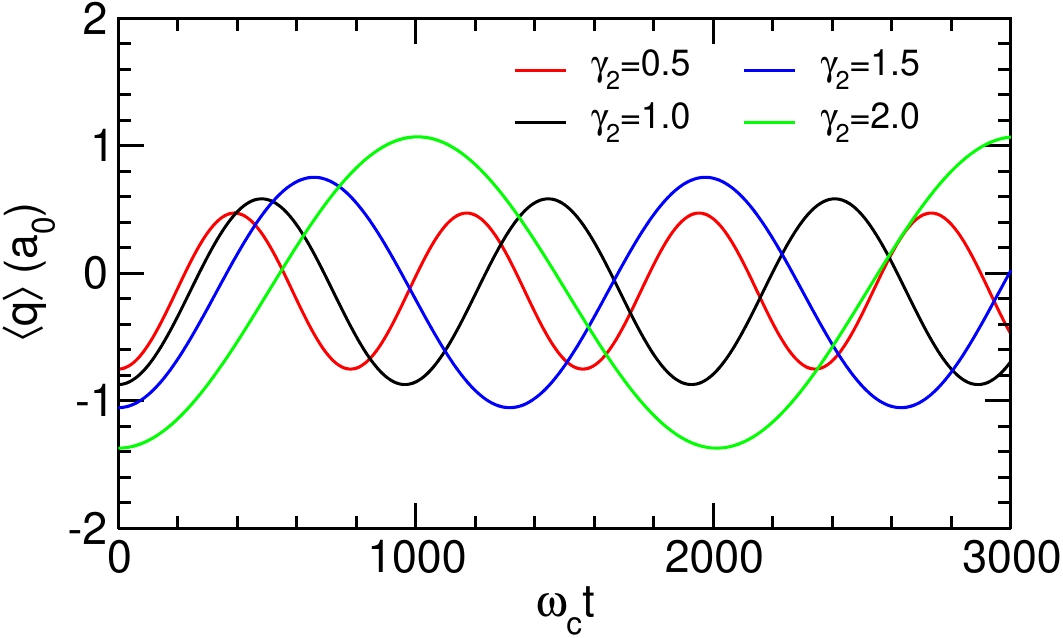}}
 \caption{Projected position $\langle q\rangle$ as a function of time,
   as obtained for $\lambda_c = \lambda_R$ assuming the initial
   wavepacket $ | \Psi (t=0) \rangle$ given in Eq.~(\ref{Eq:WP}), for
   different values of $\gamma_2$ given in {atomic
     units}.}
\label{Fig:figure7}  
\end{figure}

The TDSE was numerically solved using this initial condition for
{$\gamma_2=1.0$~au} as well as other values of $\gamma_2$
ranging from {half to twice this value,} with
$\lambda_c$ being adjusted to the new $\lambda_R$ according to the
values given in Table~\ref{tab:table3}. $\gamma_2$ is found to have
only little influence on the spectral properties of the polaritonic
system, therefore the above discussion concerning the preparation of
the initial state remains valid for other values of this
parameter. The resulting time variations of the projected position
$\langle q\rangle$ are represented in Fig.~\ref{Fig:figure7}. Since
only two polaritonic states are involved in the initial wavepacket,
the time evolution of the system follows a Rabi oscillation.
Coherent switching of the wavepacket between the two wells is clearly
visible, with a period of about {920 to 2290 $\omega_c t$, \textit{i.e.} 8 to 20 ps,} depending on the value of
$\gamma_2$. The period and the amplitude of the oscillation between
the two potential wells both increase with this parameter, as a consequence of the decrease of the energy gap $\Delta \varepsilon_R$
between the two polaritonic eigenstates (see Table~\ref{tab:table3}).

For a molecule having different reactivity properties in the left and
right wells, the preparation of such a superposition of polaritonic
states near an avoided crossing could be efficiently used to control
the outcome of the reaction. However, to exploit such features
experiments would have to be performed at low temperatures and would
require damping processes to be minimized on the picosecond time
scale, including photon loss inside the cavity and radiative or
non-radiative relaxation processes of the molecules.

\subsection{Influence of the potential asymmetry}

\begin{figure}[t]
  \centerline{\includegraphics[width=8.3cm]{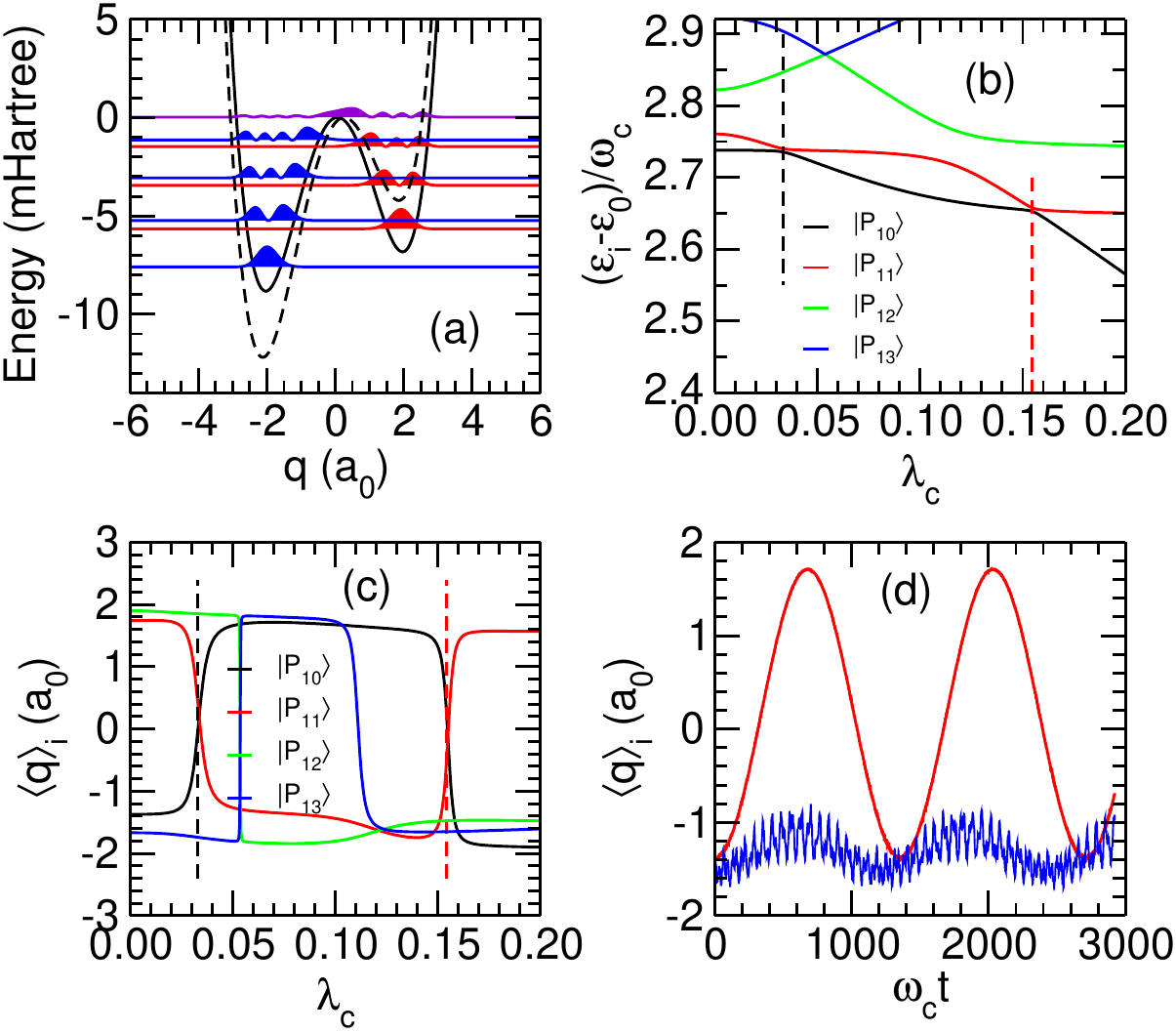}}
  \caption{Results obtained for the alternative double-well
    potential. (a) Potential energy curve (solid line) and probability
    densities associated with the lowest eigenstates. For comparison,
    the previous potential curve is superimposed as a dashed line; (b)
    relative energies $\left(\varepsilon_i
    -\varepsilon_0\right)/\omega_c$ as a function of the coupling
    strength $\lambda_c$ for $| P_{10} \rangle$, $ | P_{11}\rangle$, $
    | P_{12}\rangle$ and $ | P_{13}\rangle$; (c) average position
    $\langle q \rangle_i$ as a function of $\lambda_c$ for the same
    polaritonic states; (d) time evolution of $\langle q \rangle$
    subsequent to an initial excitation of the uncoupled state
    $|3_\ell,0\rangle$ for $\lambda_c=$ 0.033 (red curve) and 0.154
    (blue curve).}
 \label{Fig:newpotential}  
\end{figure}

The robustness of the previous analysis was finally investigated by
addressing the role of the asymmetry of the double-well potential,
which is contained in parameter $\alpha_2$ of Eq
(\ref{Eq:Pot}). Leaving parameters $\alpha_0$ and $\alpha_1$
unchanged, $\alpha_2$ is divided by 4, leading to a potential with
$N_\ell =4$ and $N_r =3$ bound states in the left and right parts of
the potential, respectively. This alternative potential and the
probability densities of its first bound states obtained from solving
the Schr\"odinger equation are depicted in
Fig. \ref{Fig:newpotential}(a).
Fixing {$\gamma_2=0.5$~au} throughout this section, the eigenenergies of the polaritonic states were determined as a function of the coupling strength $\lambda_c$. Partial results represented in Fig.~\ref{Fig:newpotential}(b) highlight the polaritonic states $|  P_{10} \rangle$ and $ |P_{11}\rangle$, which undergo two avoided crossings at $\lambda_{\rm crit}^{(1)} = 0.033$ and $\lambda_{\rm crit}^{(2)} = 0.154$. As shown from the average positions $\langle q \rangle_i$ in Fig.~\ref{Fig:newpotential}(c), these avoided crossings are both connected to transitions in which the system switches between the left and right wells of the potential. 

In the limit of vanishing couplings, the polaritonic states  $|  P_{10} \rangle$ and $ |  P_{11}\rangle$ can be assigned to the $| 3_\ell,0\rangle$ and $| 1_r,1\rangle$ states, respectively. 
While in the vicinity of the two avoided crossings, $|  P_{10} \rangle$ and $ | P_{11}\rangle$ are mainly decomposed on the uncoupled states $|3_\ell,0\rangle$, $| 0_r,2\rangle$ and $|1_r,1 \rangle$. Preparing the system in the excited state $v =3_\ell$ before entering {an initially photon-less} cavity, leading to an initial state $\ket{3_\ell,0}$, can thus be expected to produce the polaritonic states involved in these avoided crossings. The weights $W_i = |\langle P_i|3_\ell,0\rangle|^2$ of this initial state on the four polaritonic states  $|P_i\rangle$, $i=10$--13 are reported in Table~\ref{tab:table4} for the two values of $\lambda_{\rm crit}$.
\begin{table}[b]
\small
  \caption{Weights $W_i$ of state $|3_\ell,0\rangle$ on the four 
  polaritonic states $|P_i\rangle$ with $i=10$--13, near the two avoided crossings. $\Sigma$ corresponds to the sum of these four weights}
  \begin{tabular*}{0.48\textwidth}{@{\extracolsep{\fill}}cccccc}
    \hline
 $\lambda_{\rm crit}$ &$W_{10}$  & $W_{11}$ & $W_{12}$ & $W_{13} $ & $\Sigma$\\
    \hline
 $0.033$  &  55.0\% & 44.3\% & --- & 0.5\% & 99.8\% \\
 $0.154$  & 66.0\% & 5.5\% & 78.1\% & 7.5\% & 97.7\% \\
    \hline 
  \end{tabular*}
   {\label{tab:table4} }
\end{table}
%
As can be seen in this table, at $\lambda_{\rm crit}^{(1)} = 0.033$ the two polaritonic states $|P_{10} \rangle$ and $ |  P_{11}\rangle$ together represent over 99\% of the $|3_\ell,0\rangle$ wavepacket, indicating that 
it can effectively be considered as a linear combination of the polaritonic states $|  P_{10} \rangle$ and $ |  P_{11}\rangle$ alone:

\begin{equation}
    \ket{\Psi (t=0)} = \ket{3_\ell,0} \approx \frac{1}{\sqrt{2}}\left( \ket{ P_{10}} + \ket{ P_{11}}\right).
\end{equation}
%
The time evolution obtained after preparing the system in this initial state at $\lambda_c=\lambda_{\rm crit}^{(1)}$ is represented in Fig.~\ref{Fig:newpotential}(d). A Rabi oscillation between the left and right wells is found at a period of about 
{$1600 \,\omega_c t$ (about 14~ps), with an amplitude that indicates} an efficient coherent switching of the polaritonic states $|P_{10} \rangle$ and $ |  P_{11}\rangle$. 

For the second avoided crossing at  $\lambda_{\rm crit}^{(2)}=0.154$, the $|3_\ell,0\rangle$ state is no longer a superposition of the polaritonic states $|P_{10}\rangle$ and $|P_{11}\rangle$ as it now mainly involves  $\ket{P_{12}}$ and has a non negligible weight on $\ket{P_{13}}$.
In the limit of vanishing couplings, the polaritonic states  $|  P_{12} \rangle$ and $ |  P_{13}\rangle$ can be assigned to the $| 0_r,2\rangle$ and  $| 2_\ell,1\rangle$ states, respectively. However, near $\lambda_{\rm crit}^{(2)}$ the energy of $|P_{12}\rangle$ approaches the energy of the initially prepared $|3_\ell,0\rangle$ state, therefore explaining the large weight of this state on $|P_{12}\rangle$. 
The average positions of $|  P_{12} \rangle$ and $ |  P_{13}\rangle$ displayed in Fig.~\ref{Fig:newpotential}(c) show that both polaritonic states reside in the left well near $\lambda_{\rm crit}^{(2)}$ ($\langle q\rangle_i<0$ for $i = 12$, 13).
Hence it is not surprising that no significant switching between the wells is found near this avoided crossing, as confirmed by the temporal evolution of $\langle q\rangle(t)$ in Fig. \ref{Fig:newpotential}(d).

\section{Conclusion and perspectives}
\label{sec:Conclusion}

The strong interaction between a {one-dimensional potential mimicking a} bistable molecular system and a single-mode Fabry-Perot microcavity, leading to the formation of quantum polaritonic states, was computationally investigated using simple double-well potential and dipole moment models. {Although largely phenomenological in nature, the present model aims to describe a flexible molecule or a reactive system undergoing intramolecular rearrangement such as proton transfer and is rather generic in this respect.}

Considering a cavity in resonance with the first vibrational transition, a rich phenomenology of polaritonic states was {suggested} by tuning the light-matter coupling strength around polaritonic avoided crossings, {thereby exploring the idea of vacuum-modified chemistry}. The possibility of coherently switching the molecular system between the two potential wells near avoided crossings that involve polaritonic states located in either well of the potential was theoretically demonstrated.


Two possible ways of inducing coherent state switching were considered in our analysis. First with the preparation of the free molecule in a specific vibrational state before its insertion inside a cavity not containing any photon, as it could be achieved using a molecular beam perpendicular to the cavity. 
Alternatively, and assuming that (non-interacting) molecules are at thermal equilibrium inside the cavity, a suitable excitation by a {picosecond} laser of a wavepacket mixing two polaritonic states could also lead to coherent switching and produce a Rabi-like oscillation between the left and right wells. The robustness of these results with respect to the potential asymmetry and dipole parameters were both thoroughly studied.

{The switching time was found to depend significantly on the details of the system, including the potential itself but also the dipole moment function. Although the focus was on non-polar states, the dipole moment function plays a key role in polaritonic chemistry\cite{Triana2020,Triana2021} and it would be worth extending the model to the cases where either of the states, if not both, are polar. More generally, the model could be made more realistic with the goal of opening more perspectives toward the measurement of some of the predicted effects and contribute to addressing the somewhat controversial field of polaritonic chemistry.} One natural extension would be to focus on polyatomic molecules, starting with the triatomic case, and explore the possibility of transferring energy across vibrational modes 
by selectively playing on some polaritonic avoided crossings. Since the reactivity of polyatomic molecules {might} be enhanced when specific vibrational modes are excited, the coupling of the molecule with the confined light of the microcavity could lead to controlled reactivity.


\section*{Author Contributions}

Lo\"ise Attal: 
Conceptualization (equal),
Investigation (supporting),
Software (supporting),
Validation (equal),
Writing -- Original Draft (lead),
Writing -- Review \& Editing (equal);
Florent Calvo: 
Conceptualization (equal),
Validation (equal),
Visualization (supporting),
Writing -- Original Draft (supporting),
Writing -- Review \& Editing (equal);
Cyril Falvo: 
Conceptualization (equal),
Validation (equal),
Writing -- Original Draft (supporting),
Writing -- Review \& Editing (equal);
Pascal Parneix: 
Conceptualization (equal),
Investigation (lead),
Software (lead),
Supervision (lead),
Validation (equal),
Visualization (lead),
Writing -- Original Draft (lead),
Writing -- Review \& Editing (equal).

\section*{Conflicts of interest}
There are no conflicts to declare.

\section*{Acknowledgements}
The authors acknowledge financial support from GDR EMIE 3533. L.A. acknowledges financial support from MESRI through a PhD fellowship granted by the EDOM doctoral school.

\bibliography{biblio_polariton} 
\bibliographystyle{rsc} 

\end{document}